\documentclass{article}
\usepackage{graphics}
\usepackage{epsfig}
\usepackage{epstopdf}
\usepackage{amsmath}
\usepackage{float}
\usepackage{multirow}
\begin{document}
	\begin{center}
		{\Large{ Holographic Ricci DE as running vacuum with nonlinear interactions}} \\[0.2in]
		Paxy George\\
		Department of Physics, \\
		Cochin University of Science and Technology, Kochi-22, India.
	\end{center}
	\abstract
	{The holographic Ricci dark energy can be treated as a running vacuum due to its analogy in the energy density, which is a combination of H and $\dot{H}$, the model can predict either eternal acceleration or eternal deceleration. In the earlier works, we have shown that the presence of additive constant in the energy density or by considering possible interaction between dark sectors through a phenomenological term, the model can predict a transition from a prior decelerated to a late accelerated epoch. This paper analyses the cosmic evolution of holographic Ricci dark energy as running vacuum with a nonlinear interaction between dark sectors in a flat FLRW universe. We consider three possible nonlinear interaction forms which give analytically feasible solutions. We have constrained the model using the Type1a Supernova(Pantheon)+CMB(Planck 2018)+BAO(SDSS) data and evaluated the best-estimated values of all the model parameters. We have analyzed the evolution of the Hubble parameter and deceleration parameter of all three cases. We perform state finder analysis of the model, which implies the quintessence nature of the model and found that it is distinguishably different from the standard $\Lambda$CDM model. The dynamical system analysis of all three cases confirms the evolution of the universe from an unstable prior matter-dominated era to a stable end de Sitter phase.}
	\section{Introduction}
	The observational evidences from Type1a supernova (SN1a)\cite{Perlmutter:1998np,Riess:1998cb}, large scale structure (LSS)\cite{Tegmark:2003ud,Abazajian:2004aja}, and cosmic microwave background (CMB)\cite{Peiris:2003ff,Spergel:2003cb} have converged towards one of the focus of attention of modern cosmology- the accelerated expansion of the universe. An exotic component called dark energy (DE) with negative pressure was considered as the mysterious driving force responsible for the accelerated expansion of the universe. The most prominent and simplest model that explains the cosmic acceleration is the concordance model (or $\Lambda$CDM model) that considers the cosmological constant as the dark energy. The $\Lambda$CDM model describes a universe filled with cosmic components such as cold dark matter (CDM) and the cosmological constant($\Lambda$); the model accurately fits the current observational data. This model, however, was plagued with two major issues such as (a) coincidence problem, which is the fact that the present value of the matter density and the vacuum energy (VE) density is of the same order, even though the former is a rapidly decreasing function of time while the latter is just stationary, and (b) fine-tuning problem for which the observed value of the VE density is 120 orders of magnitude below the value computed using the quantum field theory (QFT). These drawbacks can be alleviated by assuming that CC can be a dynamical quantity. The main objective of these classes of models is to consider that the primary driving force that accelerates the universe is dynamical VE and, therefore, a time variable CC. The possibility of a dynamical DE is not excluded by current observations\cite{Spergel:2006hy, Davis:2007na, Komatsu:2008hk}. The recent Planck observations indicate that the DE equation of state (EOS) parameter $\omega \equiv P_{DE}/\rho_{DE}$ is close to $-1$ i.e., $\omega=-1.03\pm0.03$\cite{Aghanim:2018eyx}. Nowadays, models with time-dependent VE density $\Lambda(t)$, also known as running vacuum energy (RVE) density attains much attention. 
	%The RVE models can be accommodated by observations may appear with a non-trivial “effective EOS”,$\omega = \omega(t)$.
	The primary feature in this ideology is that the equation of state parameter of the VE is strictly equal to -1\cite{Sola:2005et, Basilakos:2009wi}.

	The idea of RVE in an expanding universe was first proposed in the literature\cite{PhysRevD.25.1019, Buchbinder:1985ew}.
	It is pretty natural to expect that the VE is a dynamical quantity as the universe expands and thereby sensitive to time-evolving functions such as the Hubble rate H = H(t). In these models, without invoking the need for scalars, a phenomenologically viable description of the dynamical nature of the VE can be achieved. By combining the phenomenological approach with the renormalization group (RG) techniques of quantum field theory (QFT) in curved spacetime, a large class of dynamical
	$\Lambda(H,\dot{H})$ models described by an even power series expansion of the Hubble rate and its derivatives were proposed by J.Sola and Shapiro\cite{Sola:2007sv, Shapiro:2000dz, Basilakos:2009wi}. However, for the present universe, only the terms $\dot{H}$ and $H^2$ can be relevant, together with an additive constant term.
	%At the present epoch,  the relevant terms can only be of order $H^2$ at most this includes $\dot{H}$. In contrast, the higher orders $H^n(n >2)$ can be used in the early universe to implement inflation successfully. These models predict VE density, which possesses a constant EOS, but their density varies as the universe expands. Nowadays, these models are exciting as it predicts a high value of vacuum energy density in the early stage of the universe to drive inflation and decay to a small value, as observed today as the universe expands.
	%(Shapiro & Sol`a2002, Sol`a 2008, Shapiro & Sol`a 2009).
	 Recent studies remarkably shows that these models that include the additive term plus one or both of the dynamical components $\dot{H}$ and $H^2$ appeared to be more favoured than the $\Lambda$CDM\cite{Basilakos:2010rs, Grande:2011xf, Basilakos:2012ra, Fabris:2006gt, Rezaei:2019xwo, Fritzsch:2016ewd}. The authors in the reference\cite{Sola:2016jky} shows that these models can provide an excellent fit to the main cosmological data SN1a+H(z)+BAO+LSS+CMB+BBN  whereas  $\Lambda$CDM model that supports a rigid cosmological constant is currently disfavored at 4.2$\sigma$. Phenomenologically, this model has a formal analogy to the holographic Ricci dark energy (HRDE), where a combination of $\dot{H}$ and $H^2$ is present on its energy density. 
	
	The holographic DE scenario is another fascinating approach to explain the late-time accelerated expansion of the universe within the context of quantum gravity. Inspired by applying the holographic principle to the universe, Li\cite{Li:2004rb, Hsu:2004ri} first introduces holographic DE. The holographic DE density is inversely proportional to the square of an appropriate length of the system that characterizes the size of the universe. One of the natural choices of this length is the inverse of the Hubble rate. However, this choice does not induce acceleration in a homogeneous and isotropic universe\cite{BouhmadiLopez:2011xi, Saridakis:2007cy}. Another choice for the length was suggested by Gao et al.\cite{Gao:2007ep, Nojiri:2005pu} in which the IR cut-off was taken to be proportional to the Ricci scalar curvature, leading to the introduction of the new holographic DE, the holographic Ricci DE. This model can derive an accelerated universe that solves the cosmological constant problem and fairly fits the observational data well. This model has been studied widely in many literature\cite{Huang:2004ai,Jamil:2009ia,Chen:2006qy,Feng:2008kz,Chang:2005ph}. The analogy between the energy density of this model with the running vacuum model motivates us to consider HRDE as a running vacuum.
	%In this paper, we focus on a dynamical model of the VE inspired by QFT in curved spacetime, namely on those based on the %structure $\Lambda(H,\dot{H})$ which is well-motivated for a Friedmann-Lematre-Robertson-Walker (FLRW)-like expanding
	%universe characterized by the Hubble rate \cite{Gomez-Valent:2014rxa,Sola:2014tta}.
	%The general of this running vacuum model is analogous with Holographic Ricci DE(HRDE), which is inspired by applying the %holograhic principle to cosmology. The holographic energy density is inversely proportional to the square of the appropriate length scale % characterizing the universe, which can be represented by a combination of $H^2$ and $\dot{H}$\cite{Gao:2007ep,FENG2008231}.
	
	In the previous works\cite{George:2015lok, George:2018myt}, we have shown that considering HRDE as a running vacuum can satisfy a conventional evolutionary status with a prior deceleration epoch followed by an accelerated epoch. In the reference\cite{George:2015lok}, we have shown that a bare cosmological constant in the RVE density is essential for the transition
	from a prior decelerated epoch to a late accelerated epoch. Later, in the work\cite{George:2018myt}, we have shown that, by considering the interaction between the RVE density and the dark matter through the phenomenological term $Q=3bH\rho_{m}$, the model can predict a transition from the prior decelerated to a late accelerated phase even without an additive constant in the VE density. The possibility to have a nonlinear interaction (NLI) between dark sectors have been discussed in many literature\cite{Oliveros:2014kla,Bolotin_2015,Arevalo:2011hh,Baldi:2010vv}. It is of great interest to study the effects of nonlinear interaction between DE and DM on the evolution of the universe, especially the influence of different forms of nonlinear interaction terms on cosmological parameters. The authors in the reference\cite{He:2008tn} have shown that nonlinear interaction
	between dark sectors is more compatible with recent observational data than linear interaction choices. In this paper, we have considered three different forms for the nonlinear interaction terms, $Q$, which allow us to find analytically feasible solutions. The paper is organized as follows. Section 2 discusses a brief introduction of the RVE model and HRDE model. This section also consists of our analysis of the evolution of the Hubble parameter corresponding to three nonlinear interaction forms. In section 3, we study the evolution of the deceleration parameter, and these models are compared with the standard $\Lambda$CDM model by performing the state finder analysis of three cases. Finally, the stability of all the nonlinear interaction terms was checked by performing dynamical system analysis that predicts an early matter-dominated and late accelerated epoch.

	\section{The model : Holographic Ricci DE as running vacuum with nonlinear interactions}
	%The running of VE density from the quantum effects of the matter fields is associated with changing the spacetime curvature.
	The running vacuum model proposed in the literature\cite{Sola:2011qr, Shapiro:2000dz, Sola:2007sv}  is based on the possibility that quantum effects in curved space time\cite{Parker:2009uva} can be responsible for the renormalization group(RG) running of the VE, which is mainly aimed to alleviate the cosmological constant problem\cite{Lima:1995ea}. The corresponding VE density, $\rho_{\Lambda}$, should be time-dependent quantities in cosmology.
	The general form of RVE density for the current universe can be represented as a function of Hubble parameter and its derivatives\cite{Sola:2014tta},
	\begin{equation}\label{rev1}
	\rho_{\Lambda}(H;\nu,\alpha)=\frac{3}{8\pi G}\left(c_0+\nu H^2+\frac{2}{3}\alpha \dot{H}+\mathcal{O}(H^4)\right),
	\end{equation}
	where, $c_0$ is an integration constant, $\nu$ and $\alpha$ are dimensionless coefficients. The values of $\nu$ and $\alpha$ can be related to the $\beta$-functions of the running vacuum and hence are assumed to be naturally small in the order of $10^{-3}$. It is possible to recover $\Lambda$CDM model for $\nu=\alpha=0$  but here we consider them as free parameters  hence we shall determine its values by fitting the model to observations.The term $\mathcal{O}(H^4)$ represent higher powers of the Hubble rate which are relevant to 
	describe inflation\cite{Lima:2012mu,Lima:2014hia,Lima:2015mca,Sola:2015rra}, but these terms has no significance for present universe and hence ignored. From the expression(\ref{rev1}), it is clear that the VE density is slowly evolving with the Hubble function $H$ and its time derivative as the universe expands. 
	
	The holographic principle is believed to be a fundamental principle of quantum gravity. Motivated by holographic principle, Cohen et al.\cite{Cohen:1998zx} 
	suggest that the quantum zero-point energy of a system with size $L$ should not exceed the mass of a black hole with the same size $L^2\rho_{\Lambda}\leq L M_p^2$. This relation relates the UV cut-off of a system to its IR cut-off. The largest $L$ saturating this inequality will give rise to a holographic energy density, 
	\begin{equation}
	\rho_{\Lambda}=3c^2M_p^2L^2,
	\end{equation}
	where c is a dimensionless constant and $M_p^2=\frac{1}{8\pi G}$ is the reduced Planck mass. Gao et al\cite{PhysRevD.79.043511} have proposed a holographic DE which is proportional to the inverse of the Ricci scalar curvature, which is aimed to avoid causality problems and is phenomenologically viable. The holographic DE proportional to the Ricci scalar for a flat FLRW universe is,
	\begin{equation}\label{ricci}
	\rho_{de}(H,\dot{H})=3\beta M_p^2(\dot{H}+2H^2),
	\end{equation}
	where $\beta$ is a dimensionless coefficient determined by observations.
	The major drawback of this model is that it will not allow a transition from a decelerated epoch to a late accelerating epoch. To alleviate this issue, a bare cosmological constant has been added to the initial form of the HRDE density, and then it resembles the general form of RVE density as given in equation(\ref{rev1}). This motivates us to consider HRDE as a running vacuum in our work. The model parameter $\beta$ can be related to the terms of the original RG running parameter as $\beta\sim \frac{\nu}{2} \sim \frac{2\alpha}{3}$\cite{George:2018myt}. In the work\cite{George:2018myt} we have shown that instead of considering an additive constant in the energy density of HRDE, it is possible to attain a transition from decelerated to accelerated epoch by accounting for a possible interaction between dark sectors.  
	The evolution equation for a spatially flat Friedmann universe is,
	\begin{equation}\label{fried}
	3H^2=\rho_m+\rho_{de},
	\end{equation}
	where $H=\frac{\dot{a}}{a},$ is the Hubble parameter, a is the scale factor, $\rho_m$ is the matter density and $\rho_{de}$ is the DE density. Here we also suppose that DM is pressureless and DE satisfies the equation of state,$\omega=-1$ which gives $P_{de}=-\rho_{de}$.  When a possible interaction between dark sectors are taken into account, the conservation equation takes the form,
	\begin{equation}\label{conser}
	\begin{split}
	\dot{\rho_m}+3H(\rho_m+P_m)=Q,\\
	\dot{\rho_{de}}+3H(\rho_{de}+P_{de})=-Q,
	\end{split}
	\end{equation} 
	where Q is the interaction term which gives the rate of energy exchange between dark sectors.  
	%For non relativistic matter,$P_m=0$ and for running vacuum $p_{de}=-\rho_{de}$. 
	The sign of Q determines the direction of flow of energy. If Q is positive the DE decays to DM while if Q is negative DM decays to DE. The interaction terms are introduced  by phenomenological considerations to avoid the coincidence problem\cite{Bolotin_2015}. The interaction term  proportional to the DE density or to the DM density or to a linear combination of both densities was well studied in literature\cite{PhysRevD.78.021302,Pan:2012ki,PhysRevD.79.043517,Wang_2006}. But in some literature\cite{Arevalo:2011hh,Baldi:2010vv,Ma:2009uw,Mangano:2002gg,Oliveros:2014kla,Sadri:2018rcp} the non linear interaction i.e., an interaction proportional to the product of DM density and DE energy density  were studied and found to be favored observationally than linear interaction between dark sectors\cite{He:2008tn}. The general form for the nonlinear interaction can be written as\cite{Arevalo:2011hh},
	\begin{equation}
	Q=3bH\rho^{m+s}\rho_m^n\rho_{de}^{-s-n},
	\end{equation} 
	where $b$ is the coupling constant, $\rho=\rho_m+\rho_{de}$ and the powers $m,n$ and $s$ specify interaction. We have analyzed the case $Q=3bH\rho_m$  for $(m,n,s)=(1,1,-1)$ in reference\cite{George:2018myt}. In the present work we consider three possible nonlinear forms for Q
	i.e., $Q=3bH\frac{\rho_m^2}{\rho}$ for $(m,n,s)=(1,2,-2)$, $Q=3bH\frac{\rho_{de}^2}{\rho}$ for $(m,n,s)=(1,0,-2)$ and $Q=3bH\frac{\rho_m\rho_{de}}{\rho}$ for $(m,n,s)=(1,1,-2).$ These particular combinations of the parameters$(m,n,s)$ gives rise to analytically solvable models with nonlinear interaction terms.
	%*******************************************************
	\subsection{Case 1: $Q=3bH\frac{\rho_m^2}{\rho}$}

	The model with this interaction term $Q=3bH\frac{\rho_m^2}{\rho}$ is named as nonlinear interaction case1(NLI1). The conservation equation(\ref{conser}) for this case takes the form,
	\begin{equation}\label{conserv}
	\begin{split}
	\frac{d\rho_{de}}{dx}=-b\frac{\rho_m^2}{H^2},\\
	\frac{d\rho_m}{dx}=\frac{b\rho_m^2}{H^2}-3\rho_m.
	\end{split}
	\end{equation} 
	Here we have changed the variable from t to $x=\ln a.$ Combining the equations (\ref{fried}),(\ref{conserv}) and (\ref{ricci}) we get
	\begin{equation}\label{differ1}
	H^2\left( \frac{d^2H^2}{dx^2}+3\frac{dH^2}{dx}\right) +9bH^4-6bH^2\rho_{de}+b\rho_{de}^2=0.
	\end{equation}
	The solution of the above differential equation give the Hubble parameter as
	\begin{equation}\label{hubble1}
	H^2=c_2e^{-Ax}\cosh[B(x-4c_1)]^F,
	\end{equation}
	where $c_1$ and $c_2$ are integration constants, A, B and F are given by
	\begin{equation}
	\begin{split}
	A=\frac{6(1+3b\beta(2\beta-1))}{4+9b\beta^2},\\
	B=\frac{3}{2}[\sqrt{1-2b(2-5\beta+2\beta^2)}],\\
	F=\frac{4}{4+9b\beta^2}.
	\end{split}
	\end{equation}
	The values of the integration constants $c_1$ and $c_2$ can be obtained by using the conditions
	\begin{equation}\label{initial}
	H|_{x=0}=H_0, \, \, \, \frac{dH^2}{dx}| _{x=0}=2H_0^2\left( \frac{\Omega_{de0}}{\beta}-2\right) ,
	\end{equation}
	
	where $\Omega_{de0}$ is the current value of the RVE density. The values of the coefficients are found to be,
	\begin{equation}\label{coef}
	\begin{split}
	c_1=\frac{1}{4B}\tanh^{-1}\left( \frac{-2}{BF}\left( \frac{\Omega_{de0}}{\beta}-2\right) -\frac{A}{BF}\right),\\
	c_2=\frac{H_0^2}{\cosh(4c_1B)^F}. 
	\end{split}
	\end{equation}

	\subsection{Case 2: $Q=3bH\frac{\rho_{de}^2}{\rho}$}
	
	We consider the interaction  $Q=3bH\frac{\rho_{de}^2}{\rho}$, nonlinear interaction case2 (NLI2) which reduce the conservation equation(\ref{conser}) as,
	\begin{equation}\label{conserv1}
	\begin{split}
	\frac{d\rho_{de}}{dx}=-b\frac{\rho_{de}^2}{H^2},\\
	\frac{d\rho_m}{dx}=\frac{b\rho_{de}^2}{H^2}-3\rho_m.
	\end{split}
	\end{equation}
	Using equations(\ref{fried}),(\ref{conserv1})  and (\ref{ricci}) we get a second order differential equation,
	\begin{equation}
	H^2\left( \frac{d^2H^2}{dx^2}+3\frac{dH^2}{dx}\right) +9b\beta^2\left( 4H^4+2H^2\frac{dH^2}{dx}+\frac{1}{4}(\frac{dH^2}{dx})^2\right)=0.
	\end{equation}
	The solution of the above equation is the Hubble parameter of this model NLI2
	\begin{equation}\label{hubble2}
	H^2=c_2e^{-Ax}\cosh[B(x-4c_1)]^F.
	\end{equation}
	The terms $A, B$ and $F$ are given by,
	\begin{equation}
	\begin{split}
	A=\frac{6(1+6b\beta^2)}{4+9b\beta^2},\\
	B=\frac{3}{2}\sqrt{1-4b\beta^2},\\
	F=\frac{4}{4+9b\beta^2}.
	\end{split}
	\end{equation}
	Here $c_1$ and $c_2$ are the integration constants whose values can be obtained from the conditions(\ref{initial}),
	\begin{equation}
	\begin{split}
	c_1=\frac{1}{4B}\tanh^{-1}\left(\frac{-2}{BF} \left( \frac{\Omega_{de0}}{\beta}-2\right) -\frac{A}{BF}\right),\\
	c_2=\frac{H_0^2}{\cosh(4c_1B)^F}.
	\end{split}
	\end{equation}

	\subsection{Case 3: $Q=3bH\frac{\rho_m\rho_{de}}{\rho}$}
	
	The interaction considered in this case is $Q=3bH\frac{\rho_m\rho_{de}}{\rho}$ which can be conveniently represented as nonlinear interaction case3 (NLI3).
	The conservation equation(\ref{conser}) using this type of interaction can be written as
	\begin{equation}\label{conserv2}
	\begin{split}
	\frac{d\rho_{de}}{dx}=\frac{-b\rho_{de}\rho_m}{H^2},\\
	\frac{d\rho_m}{dx}=\frac{b\rho_{de}\rho_m}{H^2}-3\rho_m.
	\end{split}
	\end{equation}
	Combining the above conservation equation with (\ref{fried}) and (\ref{ricci}) give a second order differential equation,
	\begin{equation}
	H^2\left(\frac{d^2H^2}{dx^2}+3\frac{dH^2}{dx}+9b\beta\left( 2H^2+\frac{1}{2}\frac{dH^2}{dx}\right) \right)-9b\beta^2\left( 4H^4+2H^2\frac{dH^2}{dx}+\frac{1}{4}\left( \frac{dH}{dx}\right) ^2 \right). 
	\end{equation}
	The solution of the above differential equation gives Hubble parameter for this case,
	\begin{equation}\label{hubble3}
	H^2=c_2e^{-Ax}\cosh[B(x+4c_1)]^F.
	\end{equation}
	The terms A, B and F in the Hubble parameter (\ref{hubble3}) are,
	\begin{equation}
	\begin{split}
	A=\frac{3(3b\beta(4\beta-1)-2)}{9b\beta^2-4},\\
	B=\frac{3}{4}\sqrt{4+9b^2\beta^2+4b\beta(4\beta-5)},\\
	F=\frac{-4}{9b\beta^2-4}.
	\end{split}
	\end{equation}
	The integration $c_1$ and $c_2$ can be evaluated by using the conditions (\ref{initial}) as,
	\begin{equation}
	\begin{split}
	c_1=\frac{1}{4B}\tanh^{-1}\left(\frac{2}{BF} \left( \frac{\Omega_{de0}}{\beta}-2\right) +\frac{A}{BF}\right),\\
	c_2=\frac{H_0^2}{\cosh(4c_1B)^F}.
	\end{split}
	\end{equation}
	
	The values of model parameter $\beta$, interaction term $b$, $\Omega_{de0}$ and $H_0$ can be constrained by using $\chi^2$ minimization technique with the following data sets. SN1a(supernova type 1a): the Pantheon sample, composed of 1048 data points in the redshift range $0.01\leq z \leq 2.3$, obtained from the compilation of 279 SN1a discovered by the Pan-STARRS1 medium-deep survey, the distance estimates
	from the Sloan Digital Sky Survey(SDSS), Supernova Legacy Survey(SNLS) and from
	various low redshift and Hubble Space Telescope(HST) samples\cite{Scolnic:2017caz}; CMB(cosmic microwave background) data: CMB shift parameter from Planck 2018\cite{Chen:2018dbv}, $\mathcal{R}=1.7502\pm 0.0046$, the redshift at last scattering $z_{\ast}=1089.92$; BAO(baryon acoustic oscillation)data: The acoustic parameter from SDSS(Solan Digital Sky Survey), $\mathcal{A}=0.484\pm0.016$, the redshift at signature of the peak oscillation measured $z=0.35$\cite{Padmanabhan:2006cia,Blake:2011en}.  
	
	We have the distance modulus as a function of the model parameters,
	%can be obtained 
	%using the expression,
	\begin{equation}\label{equ:15}
	\mu_{t}(\beta,b,\Omega_{de0},H_0,z_{i})=m-M\\=5\log_{10}\left[\frac{d_{L}(\beta,b,\Omega_{de0},H_0,z_i)}{Mpc}\right]+25,
	\end{equation}
	%predicted by calculating the distance to the 
	%respective supernovae, which will be depending the Hubble parameter predicted by the model. For a
	% spatially flat universe, the luminosity
	where $m$ and $M$ are the apparent and absolute magnitudes of the supernovae(SN1a) respectively, $z_i$ is the red shift of the
	supernova and the luminosity distance $d_{L}$ is defined as,
	\begin{equation}\label{equ:14}
	d_{L}(\beta,b,\Omega_{de0},H_0,z_i)=c(1+z_i)\int_{0}^{z_i}\frac{dz}{H(\beta,b,\Omega_{de0},H_0,z)},
	\end{equation}
	where $H(\beta,b,H_0,z)$ is the Hubble parameter as a function of the model parameters and $c$ is the speed
	of light. Now the $\chi^{2}$ function which include theoretical and observational magnitude can be written as,
	%effectively compare thdepends on the difference between the predicted and observed magnitude over all the supernovae is given as,
	\begin{equation}\label{equ:16}
	\chi^{2}(\beta,b,\Omega_{de0},H_0)=\sum_{i=1}^{n}\frac{[\mu_{t}(\beta,b,\Omega_{de0},H_0,z_{i})-\mu_{i}]^{2}}{\sigma_{i}^{2}},
	\end{equation}
	where $\mu_{i}$ is the observational distance moduli for the
	$i^{th}$ Supernova, $\sigma_{i}^{2}$ is the standard variation of the
	observation and $n=1048,$ the total number of data points. 
	%This function can then minimized to get the best fit values of the model parameters.

	For obtaining the $\chi^2$ function, we also used 
	%We also deduced the parameters using the combined data set, consisting of the previous supernova data set, 
	%the Cosmic Microwave
	%Background (CMB) data from the WMAP 7-yr observation and the Baryon Acoustic Oscillation
	%(BAO) data from Sloan Digital Sky Survey(SDSS)\cite{Wang1}. 
	The BAO signal has been directly detected by the SDSS survey at a scale $\sim$100MPc. The BAO peak parameter value was first proposed by
	D. J. Eisenstein, et al\cite{Eisenstein:2005su} and is defined as,
	\begin{equation}\label{equ:17}
	\mathcal{A}=\frac{\sqrt{\Omega_{m}}}{h(z_{1})^{\frac{1}{3}}}\left(\frac{1}{z_{1}}\int_{0}^{z_{1}}\frac{dz}{h(z)}\right)^{\frac{2}{3}},
	\end{equation}
	Here  h(z) is the Hubble parameter, $z_{1} = 0.35$ is the red shift of
	the SDSS sample\cite{PhysRevD.69.103501}. Using SDSS data from luminous red
	galaxies survey the value of the parameter $\mathcal{A}$(for flat
	universe) is given by $\mathcal{A}=0.484\pm0.016$\cite{Blake:2011en}. The
	$\chi^{2}$ function for the BAO measurement takes the form,
	\begin{equation}\label{equ:18}
	\chi^{2}_{BAO}=\frac{(\mathcal{A}-0.484)^{2}}{(0.016)^{2}}.
	\end{equation}
	The CMB shift parameter is the first peak of CMB power
	spectrum\cite{Bond:1997wr} can be written as,
	\begin{equation}\label{equ:19}
	\mathcal{R}=\sqrt{\Omega_{m}}\int_{0}^{z_{2}}\frac{dz}{h(z)},
	\end{equation}
	Here $z_{2}$ is the redshift at the last scattering surface. 
	
	From the Planck 2018 data, $z_{2}=1089.92$. At this redshift,
	$z_{2}$, the value of shift parameter would be $\mathcal{R}=1.7502\pm
	0.0046$\cite{Chen:2018dbv}. The $\chi^{2}$ function for the CMB measurement can be
	written as,
	\begin{equation}\label{equ:20}
	\chi^{2}_{CMB}=\frac{(\mathcal{R}-1.7502)^{2}}{(0.0046)^{2}}.
	\end{equation}
	Considering three cosmological data sets together, i.e., (SN1a+BAO+CMB),  the total $\chi^{2}$ function is
	then given by,
	%\begin{equation}\label{equ:21}
	%\chi^{2}_{total}=\chi^{2}_{SNe}+\chi^{2}_{BAO}+\chi^{2}_{CMB}.
	%\end{equation}
	
	%The best estimated parameter values are obtained by performing combined $\chi^2$ minimization technique as in %references\cite{George:2015lok,George:2018myt} i.e., 
	\begin{equation}
	\chi_{total}^2(\beta,b)=\chi_{SN1a}^2+\chi_{CMB}^2+\chi_{BAO}^2. 
	\end{equation}  
	The best fit values for the model parameters corresponding to the 68.3$\%$ confidence interval for the three cases are given in Table \ref{table1}. Along with this, the best fit values of the absolute magnitude, M, of SN1a can be evaluated. From the table, it is clear that values of the model parameter $\beta$ for NLI2 and NLI3 are the same i.e.,$\beta \sim 0.456$ whereas NLI1 shows a slight change in its value. The value of the interaction term is $\sim$ 0.003 for NLI1 and NLI3, whereas for NLI2, its value is 0.001. For all the three cases, the best fit values of $\Omega_{de0}$, $H_0$, and M are found to be 0.72, 68.7, 19.39, respectively. The chi-square degrees of freedom, $\chi^2_{d.o.f}=\frac{\chi^2_{min}}{m-n}$, where $m$ is the number of data points, and $n$ is the number of the model parameter, is found to be close to one. 
	
\begin{figure}[H]
		\centering
		\includegraphics[scale=0.45]{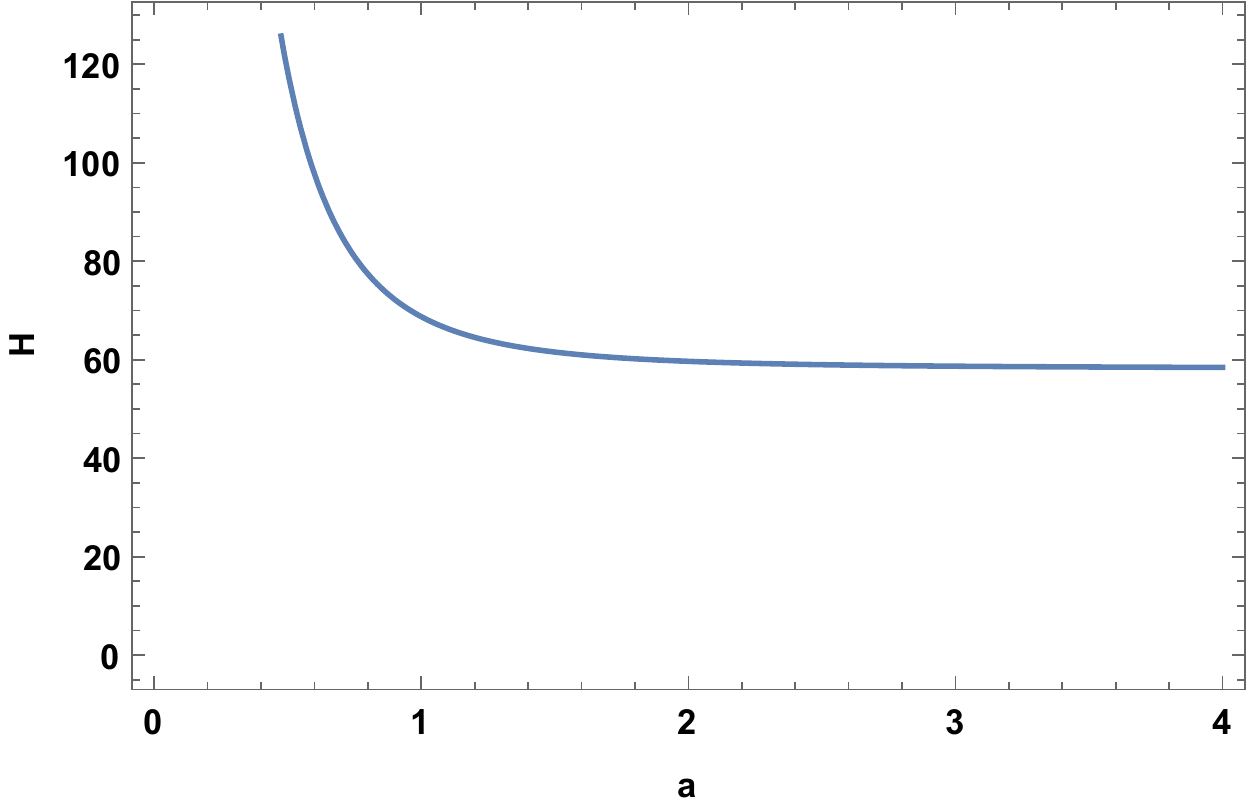}
		\caption{The plot shows the evolution of Hubble parameter for the case NLI1.}\label{fig1a}
	\end{figure}

\begin{figure}[H]
		\centering
		\includegraphics[scale=0.45]{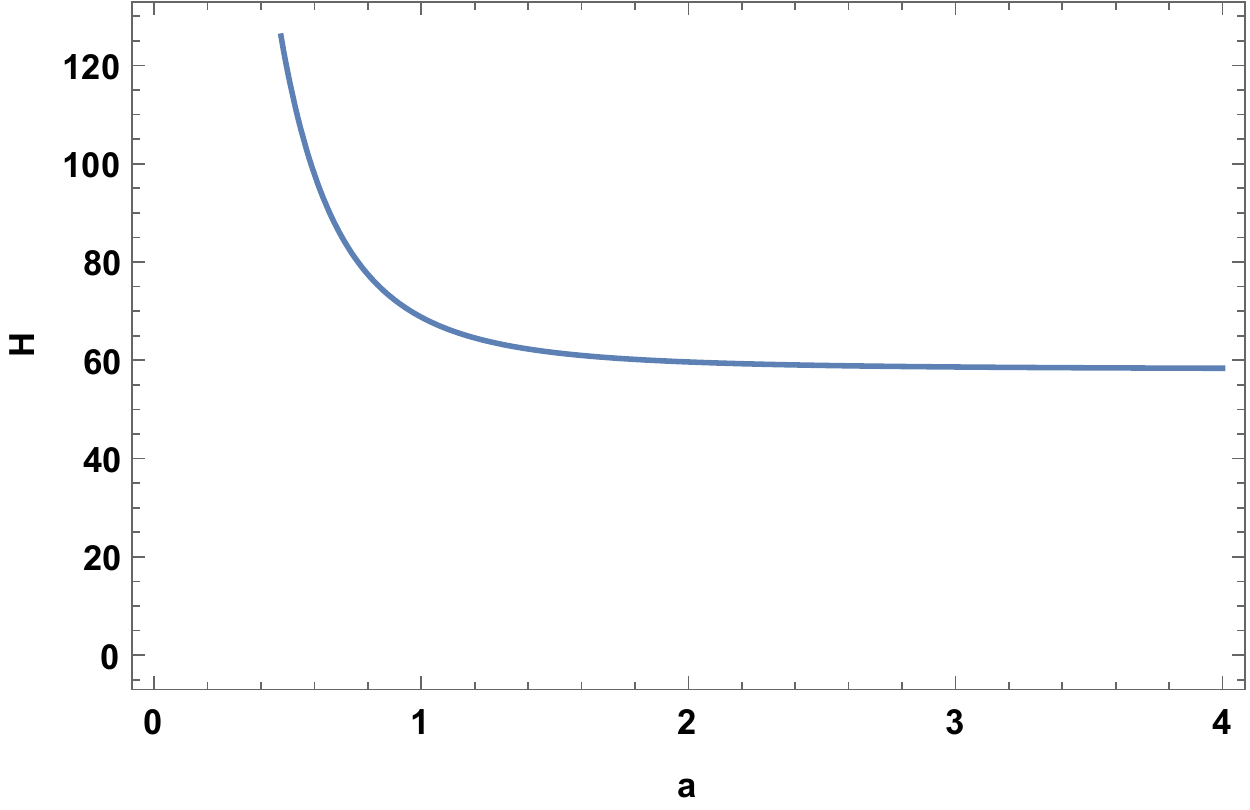}
		\caption{The plot shows the evolution of Hubble parameter for the case NLI2.}\label{fig1b}
	\end{figure}

\begin{figure}[H]
		\centering
		\includegraphics[scale=0.45]{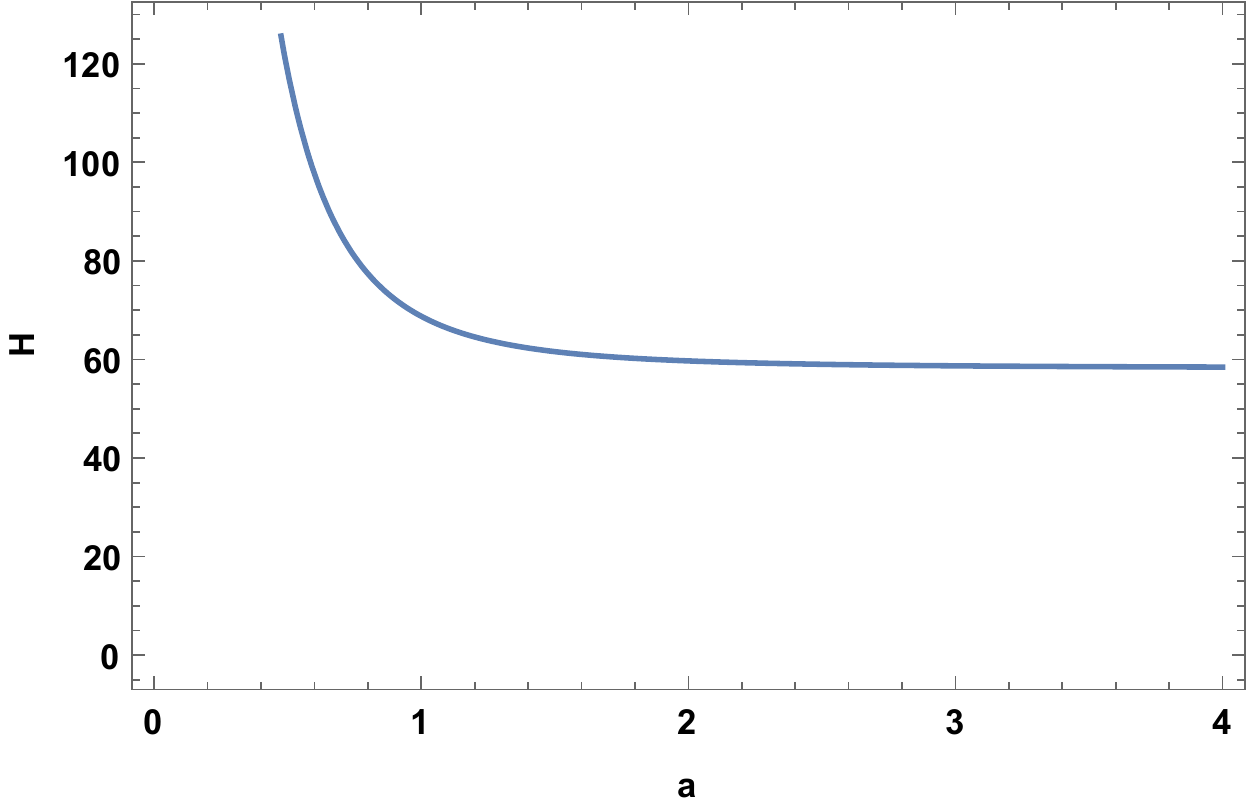}
		\caption{The plot shows the evolution of Hubble parameter for the case NLI3.}\label{fig1c}
	\end{figure}

	%\begin{columns}
	%\column{0.5\textwidth}
	\begin{table}[h]
		\renewcommand{\arraystretch}{1.5}
		\caption{The parameter values for the SN1a+CMB+BAO data.}\centering
		\small\addtolength{\tabcolsep}{5pt}
		\begin{tabular}{|c| c| c| c| }
			%	\small
			%	\begin{tabular} 
			\hline
			$Paramter$ & $NLI1$ & $NLI2$  & $NLI3$ \\ 
			%(WMAP7)         &    (WMAP7)     &   (Planck2013)    &     (Planck2013) &      \\ 
			\hline
			$\beta$ &	$0.4632_{-0.003}^{+0.003}$ &  $0.4712_{-0.002}^{+0.004}$ & $0.4656_{-0.002}^{+0.002}$ \\ \hline
			$b$    &    $0.003_{-0.006}^{+0.006}$ & $0.001_{-0.006}^{+0.005}$ & $0.003_{-0.004}^{+0.004}$ \\ \hline
			$\Omega_{de0}$ & $0.721_{-0.016}^{+0.015}$ & $0.719_{-0.016}^{+0.016}$ & $0.720_{-0.014}^{+0.013}$\\ \hline   
			$H_0$ & $68.74_{-0.42}^{+0.45}$ & $68.77_{-0.46}^{+0.47}$ & $68.76_{-0.46}^{+0.46}$\\ \hline
			M & 19.39 & 19.39 & 19.39\\ \hline
			$\chi^2_{min}$ & 1024.27 & 1024.23 & 1024.24\\ \hline
		\end{tabular} \label{table1}
		%	\end{small}
		%\end{small}
	\end{table}
	%	\column{0.5\textwidth}
	%\end{columns}
	The evolution of the Hubble parameter with scale factor for the three cases using the best fit values of the model parameters are shown in Figure (\ref{fig1a}), (\ref{fig1b}) and (\ref{fig1c}). From the plot, it is clear that the density becomes infinity at the origin, which predicts the presence of a big bang at the origin. This was again confirmed by the asymptotic behavior of the Hubble parameter of three cases given in equations (\ref{hubble1}), (\ref{hubble2}) and (\ref{hubble3}),i.e., when $a\to0$ tends H $\to$ $\infty$. 
	\section{Background analysis}
	
	\subsection{Deceleration parameter}
\begin{figure}[H]
		\centering
		\includegraphics[scale=0.45]{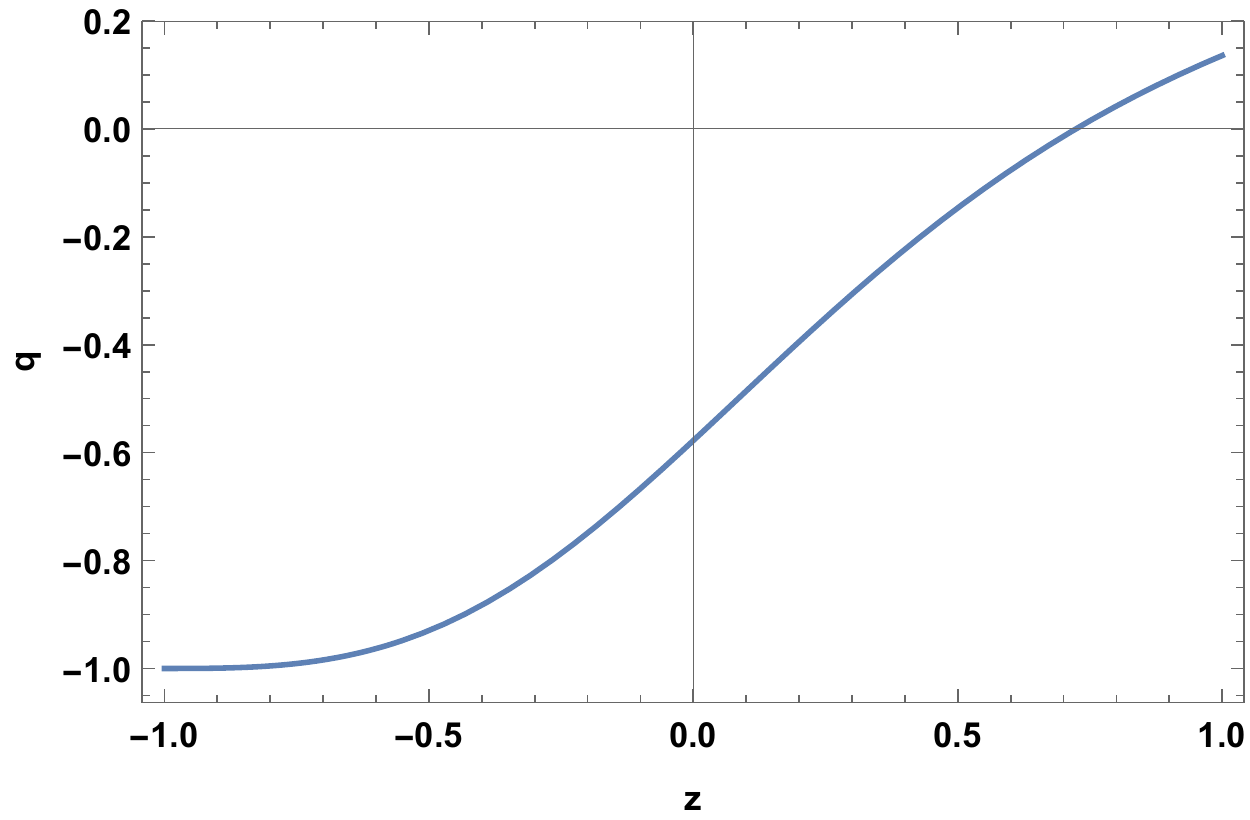}
		\caption{The q-z plot for the case NLI1.}\label{fig2a}
	\end{figure}

\begin{figure}[H]
		\centering
		\includegraphics[scale=0.45]{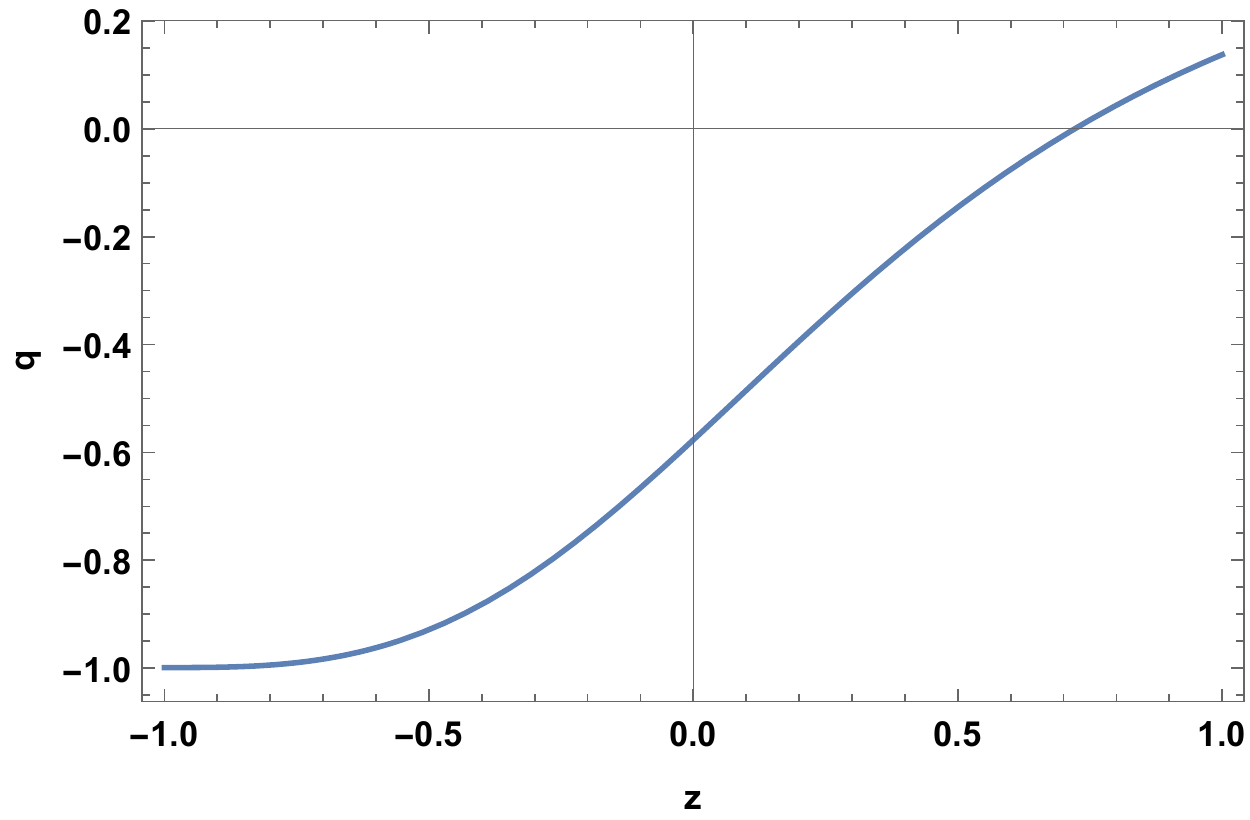}
		\caption{The q-z plot for the case NLI2}\label{fig2b}
	\end{figure}

\begin{figure}[H]
		\centering
		\includegraphics[scale=0.45]{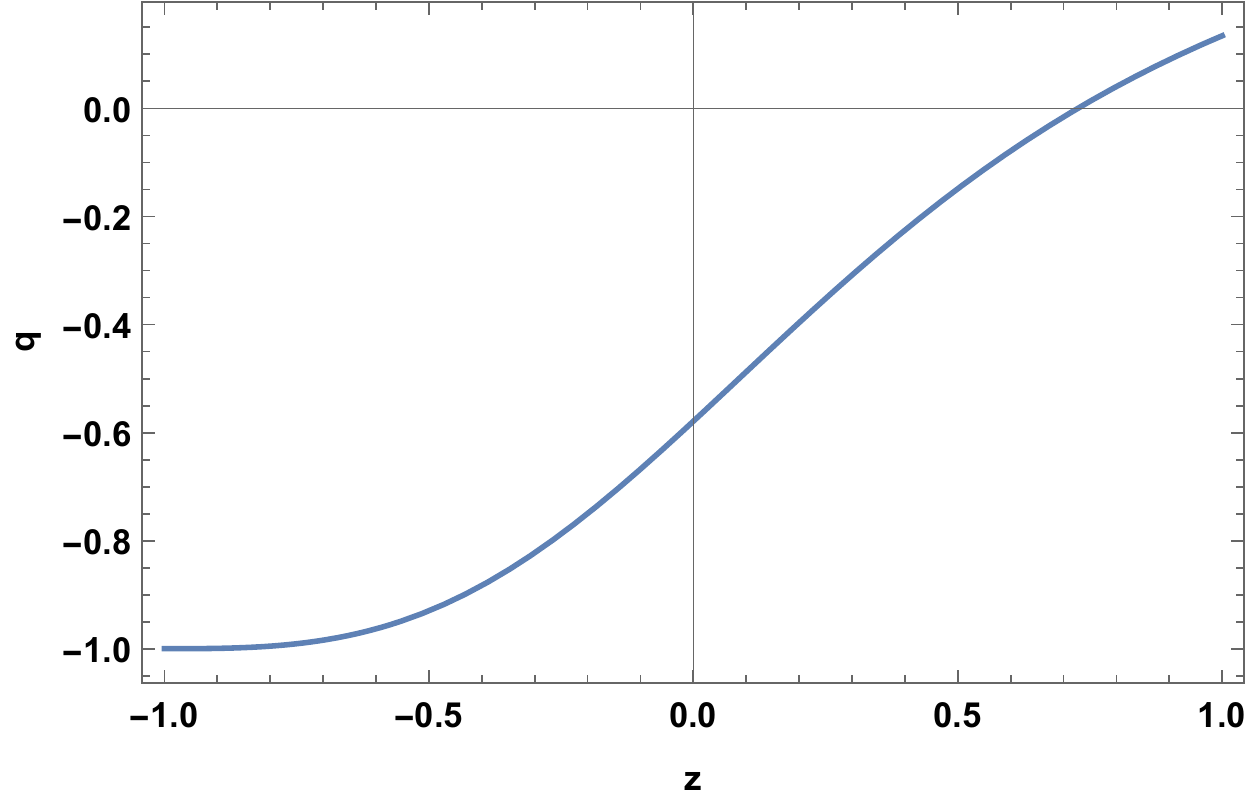}
		\caption{The q-z plot for the case NLI3}\label{fig2c}
	\end{figure}

	The decelerating/accelerating nature of the universe can be verified by studying the evolution of the deceleration parameter, which can be defined as
	\begin{equation}
	q=-1-\frac{\dot{H}}{H^2}.
	\end{equation}
	If the decelerating parameter is negative then the universe is in the accelerating phase and vice versa.
	By substituting the values of the Hubble parameter and its derivatives for three interaction cases we get,
	\begin{equation}\label{q1}
	\begin{split}
	q_{(NLI1)}=-1-\frac{-Ac_2e^{-Ax} \cosh[B(-4c_1+x)]^F + 
		B c_2 e^{-Ax}F\cosh[B(-4 c_1 + x)]^{-1 + F}Sinh[B(-4c_1+x)]}{c_2e^{-Ax} \cosh[B(x-4 c_1)]^F}.\\
	q_{(NLI2)}=-1-\frac{-Ac_2e^{-Ax}\cosh[B(-4c_1+x)]^F + 
		B c_2 e^{-Ax}F\cosh[B(-4 c_1 + x)]^{-1+F}Sinh[B(-4c_1+x)]}{c_2e^{-Ax} \cosh[B(x-4 c_1)]^F}.\\
	q_{(NLI3)}=-1-\frac{-Ac_2e^{-Ax} \cosh[B (4 c_1 + x)]^F + 
		Bc_2 e^{-Ax}F\cosh[B (4c_1 + x)]^{-1+F}Sinh[B (4 c_1+x)]}{c_2 e^{-Ax} \cosh[B (x+4c_1)]^F}.
	\end{split}
	\end{equation}
	The variation of the deceleration parameter $q$ with the red shift(z) for three cases are shown in figure (\ref{fig2a}), (\ref{fig2b}) and (\ref{fig2c}).  From the plots, the transition red shift for the three cases are found to be $z_T\sim 0.72$. The transition red shift of $\Lambda$CDM model in combination with SN1a+CMB data is reported in the range $z_T = 0.45 - 0.73$\cite{Alam:2004jy}.  The transition red shift predicted by the three cases of  present model are found to be in agreement with the upper limit of the corresponding $\Lambda$CDM range. The present value of the deceleration parameter of the models corresponding to $z=0$ are found to be $q_0\sim -0.577_{-0.037}^{+0.057}$ which is close to the WMAP value of the deceleration parameter $q_0\sim-0.60$\cite{Hinshaw:2008kr} and with the Planck(2015) value $q_0 \sim -0.55$\cite{Ade:2015xua}.  From the q-z plot, it is clear that the q-z curve stabilizes at $\sim -1$.
	
	\subsection{Statefinder analysis }
\begin{figure}[H]
		\centering
		\includegraphics[scale=0.45]{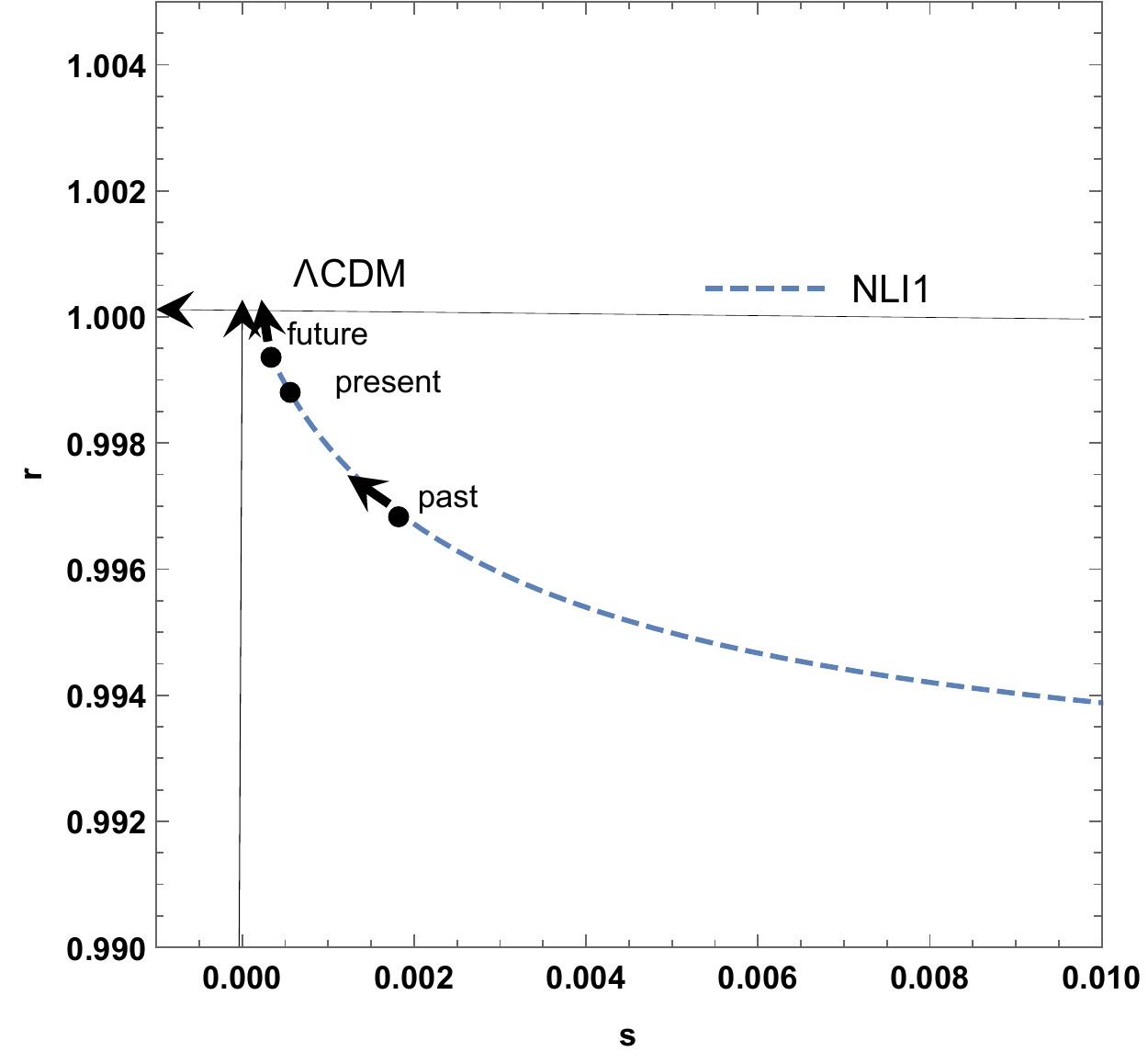}
		\caption{The r-s plot for NLI1}\label{fig3}
	\end{figure}

\begin{figure}[H]
		\centering
		\includegraphics[scale=0.45]{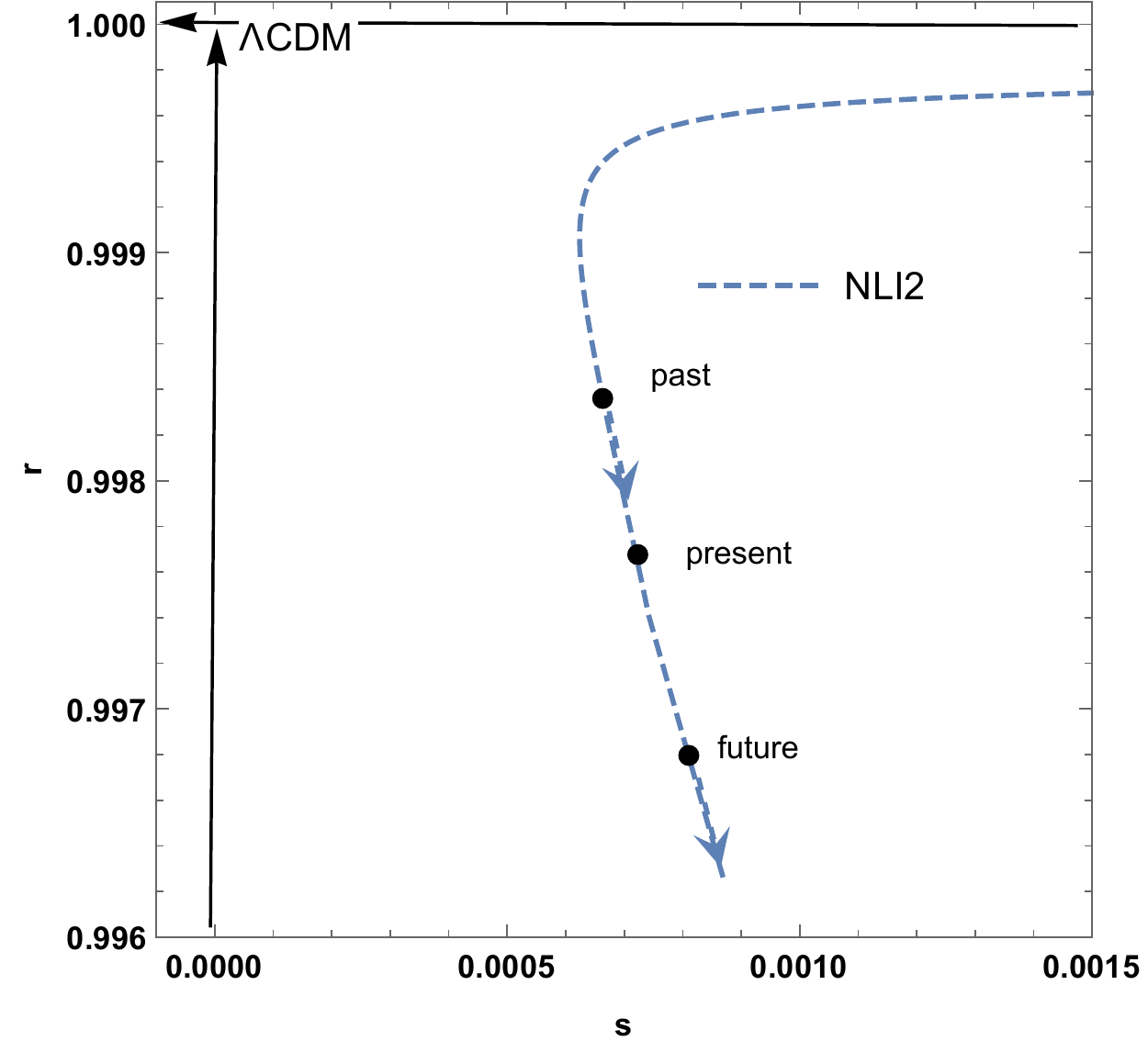}
		\caption{The r-s plot for NLI2}\label{fig4}
	\end{figure}

\begin{figure}[H]
		\centering
		\includegraphics[scale=0.45]{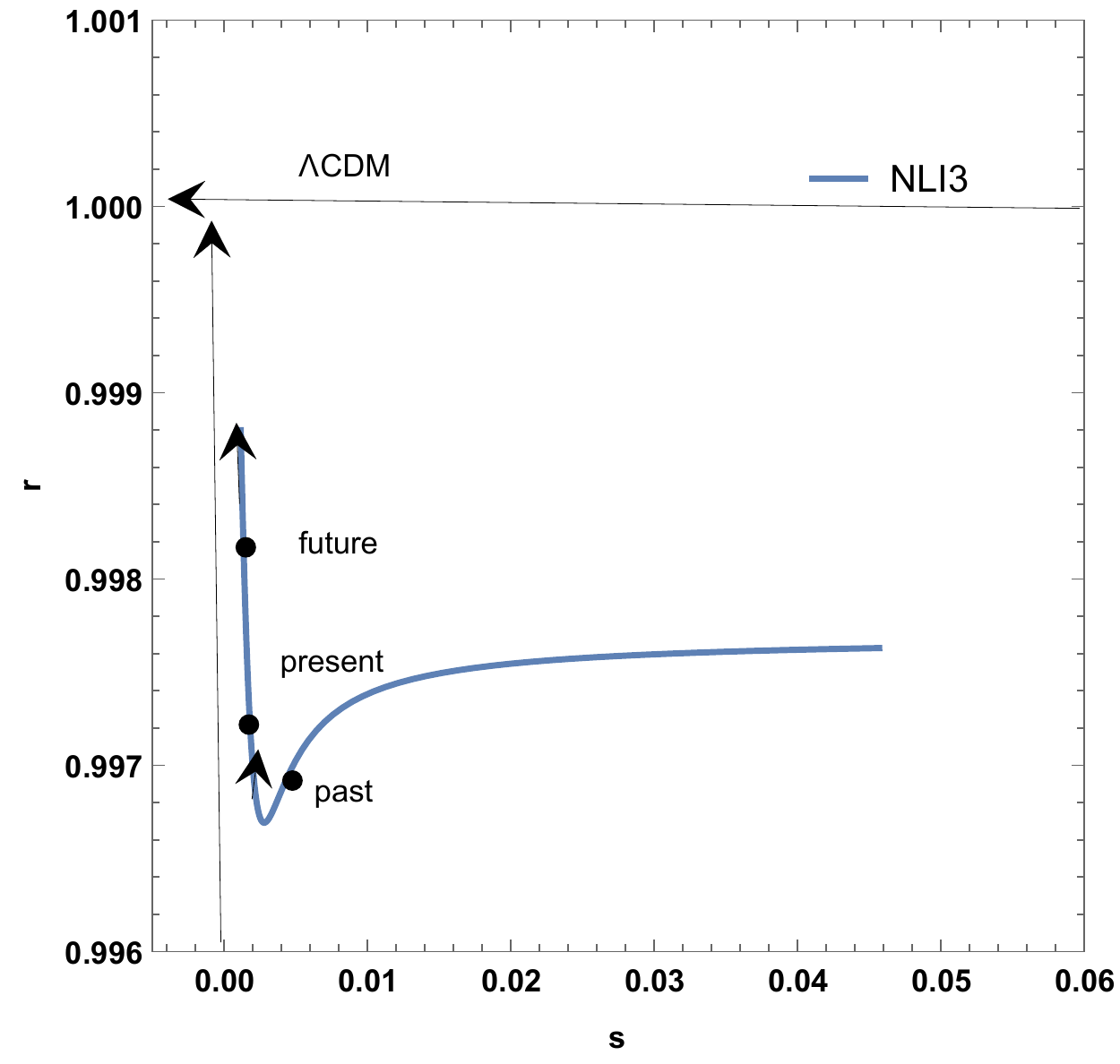}
		\caption{The r-s plot for NLI3}\label{fig5}
	\end{figure}

	Statefinder analysis is a geometric diagnostic technique for contrasting various DE models with the standard $\Lambda$CDM model introduced by Sahni et al.\cite{Sahni:2002fz}. This analysis allows one to characterize the properties of DE in a model-independent manner. Usually, Hubble parameter, scale factor, deceleration parameter, etc., are used for describing a model; it is better to use quantities involving higher derivatives
	of H or scale factor. The state finder parameter pair \{r,s\}depends on the third-order derivative of the scale factor. The trajectories in the r-s plane for various DE models are generated to bring about the discrimination between these models. For the concordance $\Lambda$CDM model, the state finder parameter is a fixed point $\{r,s\}=\{1,0\}$ but for quintessence and phantom models the trajectories lie in the region $s > 0; r < 1$ whereas for Chaplygin gas models trajectories lie in the region $s < 0; r > 1$. 
	Thus the difference in the trajectory of a given model from the standard $\Lambda$CDM model discriminates the model up to a great extent.
	%The evolution of these parameters for different DE models distinguishes the models from each other and also from the standard %$\Lambda$CDM model. 
	The state finder parameter $\{r,s\}$ can be defined as,
	
	\begin{equation}\label{state1}
	\begin{split}\\
	r=\frac{\dddot{a}}{aH^{3}}=\frac{1}{2h^{2}}\frac{d^{2}h^{2}}{dx^{2}}+\frac{3}{2h^{2}}\frac{dh^{2}}{dx}+1,\\
	s=\frac{r-1}{3(q-\frac{1}{2})}=-\frac{\frac{1}{2h^{2}}\frac{d^{2}h^{2}}{dx^{2}}+\frac{3}{2h^{2}}\frac{dh^{2}}{dx}}{\frac{3}{2h^{2}}\frac{dh^{2}}{dx}+\frac{9}{2}}.
	\end{split}
	\end{equation}
	By substituting the value of Hubble parameter and its derivatives in the above equation(\ref{state1}) for the three interaction cases, the evolution of the $\{r,s\}$ parameter was plotted and is given in the figure(\ref{fig3}), (\ref{fig4}) and (\ref{fig5}).
	
	The value of the statefinder parameter corresponding to the current epoch for the three cases is found to be
	\begin{equation}
	\begin{aligned}
	(r_0,s_0)_{NLI1}=(0.999,0.003),\qquad 
	(r_0,s_0)_{NLI2}=(0.997,0.007),\\
	(r_0,s_0)_{NLI3}=(0.997,0.008).
	\end{aligned}
	\end{equation}
	This indicates that all three cases are distinguishably different from the $\Lambda$CDM model. The present values of$\{r,s\}$ parameter of NLI2 and NLI3 are almost the same, whereas there is a slight difference in its value for the case of NLI1. The evolution of the trajectories of the three cases shows that the state finder parameter lies in the region corresponding to $r<1$ and $s>0$ until it reaches the fixed $\Lambda$CDM point resembles the quintessence nature of DE\cite{Zhang:2005rj}. The trajectory of NLI1 and NLI3 is approaching the $\Lambda$CDM fixed point in the future, whereas NLI2 is diverging away from the $\Lambda$CDM point. 
	\section{Phase Space Analysis}
	
	We study the asymptotic behavior of the present model for the three nonlinear interaction cases separately by using dynamical system tools. We define a new set of dimensionless variables,
	\begin{equation}\label{phase1}
	u=\frac{\rho_{m}}{3H^{2}} \hspace{0.5in}
	v=\frac{\rho_{de}}{3H^{2}}
	\end{equation}
	
	We formulate the autonomous coupled differential equations by using these variables, Friedmann equation and the corresponding conservation equations for each case as, 
	
	For the case NLI1:
	\begin{equation}\label{phase2}
	\begin{split}
	u'=\frac{du}{dx}=3(bu-1)u-6(1-v)\left(\frac{v}{2}-\frac{1}{2}\right)=f(u,v),\\
	v'=\frac{dv}{dx}=-3bu^2-6v\left(\frac{v}{2}-\frac{1}{2}\right)=g(u,v).
	\end{split}
	\end{equation}
	
	For the case NLI2:
	\begin{equation}\label{phase3}
	\begin{split}
	u'=\frac{du}{dx}=3(bv^2-u)-6u\left(\frac{v}{2}-\frac{1}{2}\right)=f(u,v),\\
	v'=\frac{dv}{dx}=-3bv^2-6(1-u)\left(\frac{v}{2}-\frac{1}{2}\right)=g(u,v).
	\end{split}
	\end{equation}

	For the  case NLI3:
	\begin{equation}\label{phase4}
	\begin{split}
	u'=\frac{du}{dx}=3u(bv-1)-6u\left(\frac{v}{2}-\frac{1}{2}\right)=f(u,v),\\
	v'=\frac{dv}{dx}=-3buv-6(1-u)\left(\frac{v}{2}-\frac{1}{2}\right)=g(u,v).
	\end{split}
	\end{equation}
	Here the prime refers to differentiation with the variable  $x=\ln a.$
	The equilibrium solutions are obtained by equating  $u'=0$ and $v'=0.$ The corresponding critical points for the three interaction cases are, $(u_{\ast},v_{\ast})=(0,1),(\frac{1}{1+b},\frac{b}{1+b})$ for NLI1, $(u_{\ast},v_{\ast})=(1,0),(\frac{b}{1+b},\frac{1}{1+b})$ for NLI2, $(u_{\ast},v_{\ast})=(0,1),(1,0)$ for NLI3. The state $(0,1)$ corresponds to the end de Sitter phase where as the state $(1,0)$ represents early matter-dominated phase. The state $(\frac{1}{1+b},\frac{b}{1+b})$ produce a matter-dominated epoch and that $(\frac{b}{1+b},\frac{1}{1+b})$ shows a DE-dominated era. The stability of the dynamic system in the neighborhood of the critical point can be checked by by considering small perturbations around the critical point as $u=u_{\ast}+\delta u$, $v=v_{\ast}+\delta v$.
	Linearize the system of equations (\ref{phase2}),(\ref{phase3}) and (\ref{phase4}) with respect to these perturbation gives a matrix equation,
	\begin{equation}\label{phase5}
	\begin{bmatrix}
	\delta u'\\
	\delta v'
	\end{bmatrix}
	=
	\begin{bmatrix}
	(\frac{\partial f}{\partial u})_{\ast} &(\frac{\partial f}{\partial
		v})_{\ast}\\
	(\frac{\partial g}{\partial u})_{\ast} &(\frac{\partial g}{\partial
		v})_{\ast}
	\end{bmatrix}
	\begin{bmatrix}
	\delta u\\
	\delta v
	\end{bmatrix}
	\end{equation}
	
	The Jacobian matrix i.e., $2\times2$ matrix in the right of equation(\ref{phase5}) for the three cases of the present model are,
	
	for NLI1,
	\begin{equation}\label{phase6}
	\begin{bmatrix}
	3bu_{\ast}+3(bu_\ast-1) & -3(1-v_{\ast})+3(v_{\ast}-1)\\
	-6bu_{\ast} &-3(v_{\ast}-1)-3v_{\ast}
	\end{bmatrix}
	\end{equation}
	
	for NLI2,
	\begin{equation}\label{phase7}
	\begin{bmatrix}
	-3 & -3(1-v_{\ast})+3(-1+v_{\ast})+6bv_{\ast}\\
	0  & -3(-1+v_{\ast})-3v_{\ast}-6bv_{\ast}
	\end{bmatrix}
	\end{equation}

	for NLI3,
	\begin{equation}\label{phase8}
	\begin{bmatrix}
	-3(-1+v_{\ast})+3(-1+bv_{\ast}) & -3u_{\ast}+3bu_{\ast}\\
	3(-1+v_{\ast}) &-3(1-v_{\ast})-3bu_{\ast}
	\end{bmatrix}
	\end{equation}
	Diagonalizing the above matrices gives the eigenvalues. The stability of the critical points are inferred from the sign of
	eigenvalues of the Jacobian matrix. If all the eigenvalues are positive, then the trajectories from the neighbourhood of these points diverge away from that critical point and then it is an unstable node. If all the eigenvalues are negative, the trajectories from the neighbourhood of the critical point converge to that point, and then such critical points are stable attractors. If it consists of both positive and negative eigenvalues, then the critical point is said to be a saddle point for which the trajectories are diverging or converging from these points. The critical points, eigenvalues and nature of the eigenvalues obtained for the three different interaction cases are summarized in the table(\ref{table2})  
	
	\begin{table}[ht]
		\renewcommand{\arraystretch}{2.3}
		\caption{Critical points and eigenvalue values} \centering
		\begin{tabular}{|c |c |c |c|}
			\hline
			$Models$ &$Critical points$ & $Eigen values$ & $Nature$\\ [0.5ex] \hline \hline
			
			\multirow{2}{4em}{NLI1}    & (0.997,0.003)  & (-3,3) & saddle \\
			& (0,1)  & (-3,-3) & stable\\[0.5ex] \hline
			
			\multirow{2}{4em}{NLI2}  & (1,0) & (-3,3) & saddle \\
			& (0.001,0.999)  & (-3,-3) & stable\\[0.5ex] \hline
			\multirow{2}{4em}{NLI3} &(1,0) & (-3,2.99) & saddle \\
			&(0,1) & (-3,-2.99) & stable \\[0.5ex] \hline
			
		\end{tabular}
		\label{table2}
	\end{table}
	
	For the case NLI1: the critical point $(0.979,0.021)$ is a saddle point since one of the eigenvalues is positive and the other is negative. The critical point $(0,1)$ is a stable point for which the eigenvalues are negative. For the case, NLI2:the critical point $(1,0)$ is a saddle point seeing that the eigenvalues consist of positive and negative values. The critical point $(0.072,0.928)$ is stable since the eigenvalues are negative. For the case, NLI3: the critical point $(1,0)$ is a saddle point for which the eigenvalues consist of positive and negative values. The critical point $(0,1)$ is stable since all the eigenvalues are negative. The asymptotic behaviour of all three cases predicts the transition of the universe from the early decelerated epoch to the stable later accelerated epoch.     
	
	\begin{figure}[H]
		\centering
		\includegraphics[scale=0.45]{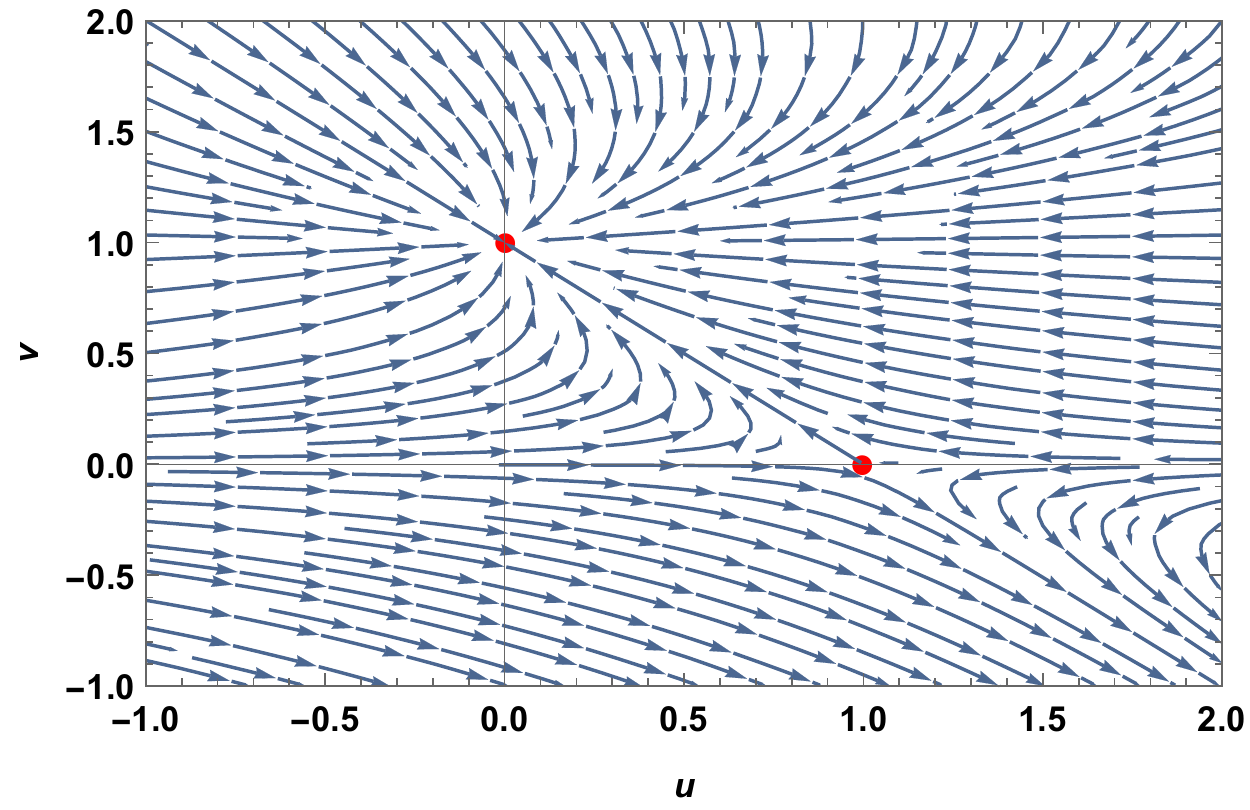}
		\caption{Phase space plot for NLI1}\label{fig6}
	\end{figure}
	
	\begin{figure}[H]
		\centering
		\includegraphics[scale=0.45]{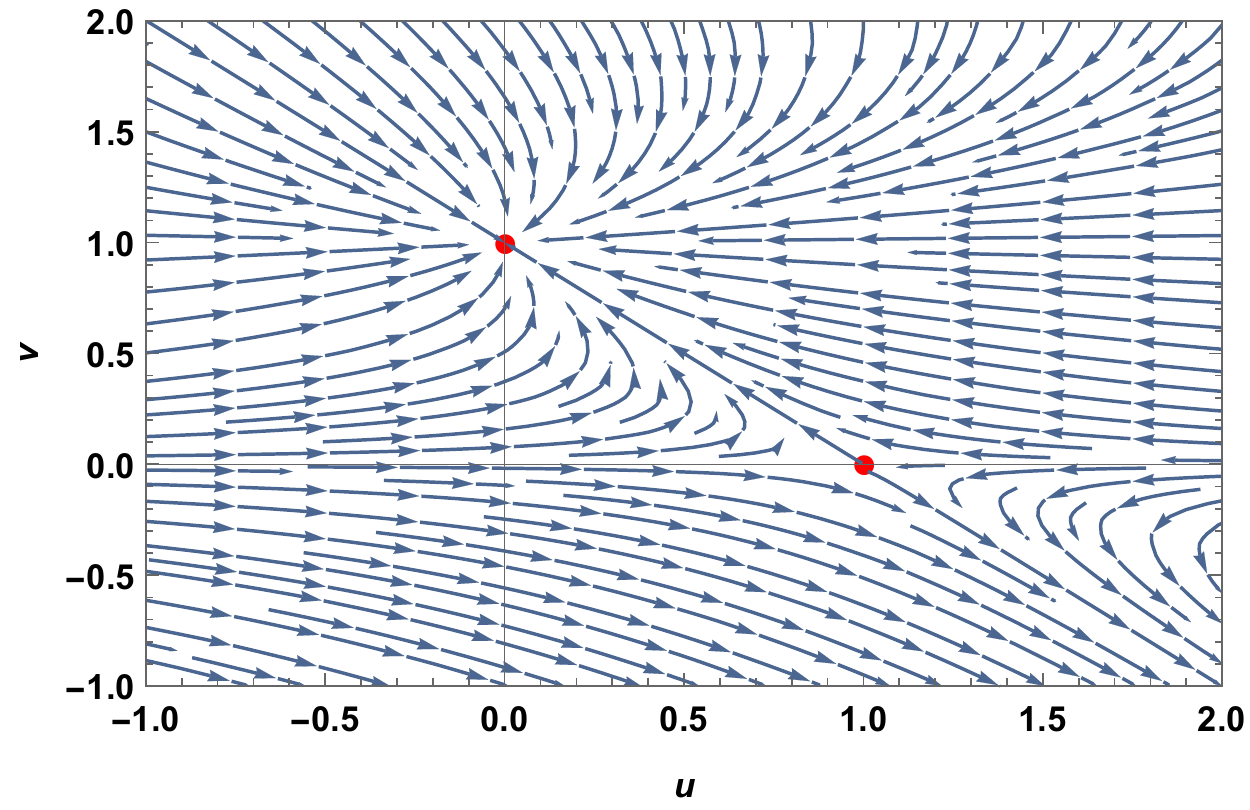}
		\caption{Phase space plot for NLI2}\label{fig7}
	\end{figure}
	
	\begin{figure}[H]
		\centering
		\includegraphics[scale=0.45]{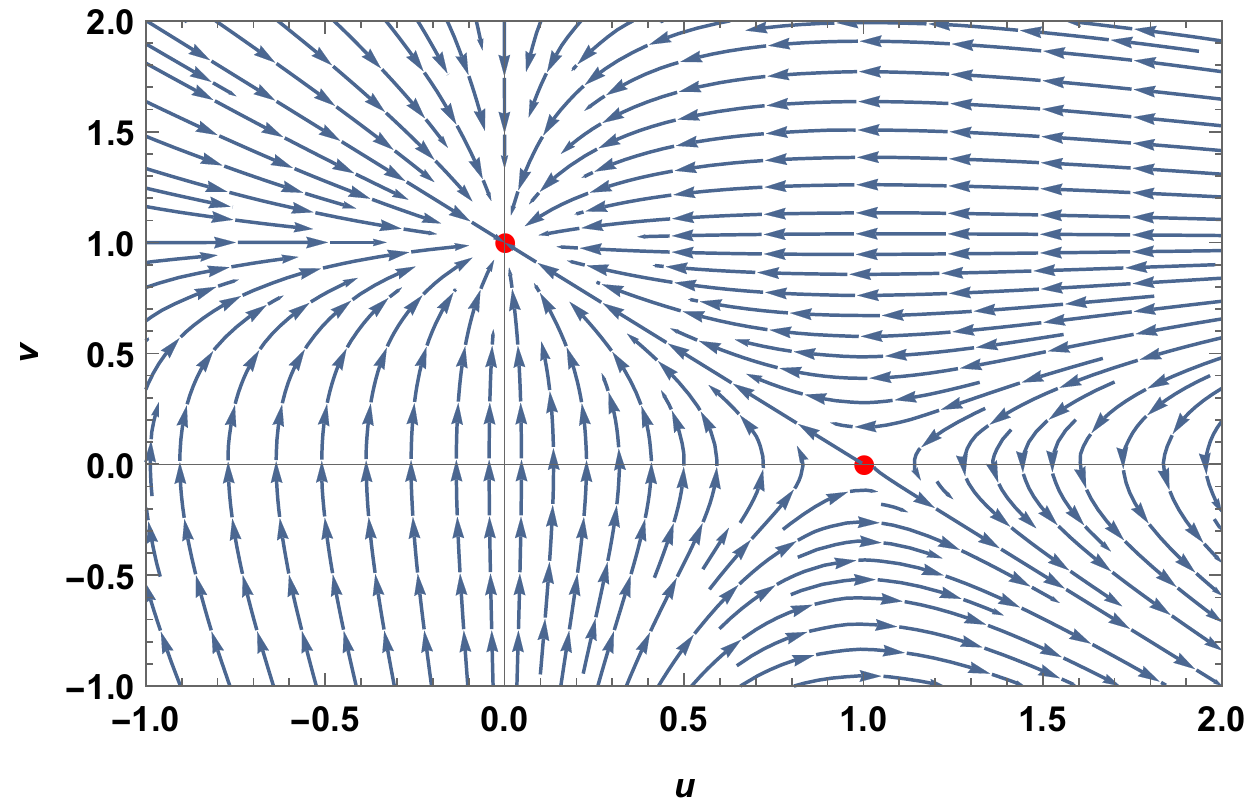}
		\caption{Phase space plot for NLI3}\label{fig8}
	\end{figure}

	\section{Conclusion}
	
	RVE model has emerged as a new look into the cosmological constant problem. These models are well motivated from the renormalization approach of quantum field theory in curved spacetime indicates the possibility of slowly varying CC\cite{Sola:2005et, Sola:2007sv, Sola:2011qr}. These models fit the current observational data quite successfully compared to the $\Lambda$CDM. The general form of RVE density consists of a combination of $H^2$ and $\dot{H}$ terms. This is analogous to the holographic Ricci DE density, where a combination of $H^2$ and $\dot{H}$ terms are also present in it. So we consider holographic Ricci DE as a running vacuum. In the earlier works, we consider holographic Ricci DE with an additive constant in its energy density. We have shown that this additive constant in the energy density is responsible for causing a transition from prior decelerated to a late accelerated phase of the universe\cite{George:2015lok}. Then we have shown that by accounting for the interaction between the dark sectors through a phenomenological term $Q=3bH\rho_{m}$ transition from decelerated to accelerated epoch is possible without an additive constant in the DE density\cite{George:2018myt}. The cosmology with the interaction between DM and DE has gained massive attention because several cosmic puzzles such as the cosmological constant problem, coincidence problem as well as some recent observational issues, such as the $H_0$ tension, $\sigma$8 tension, and the crossing of the phantom barrier without any scalar field theory can be solved by the allowance of an interaction between the dark sectors\cite{DiValentino:2019jae, Kumar:2019wfs, Pan:2020bur}. The observational data collected from various astrophysical sources have favoured interaction between the dark sectors. Since the precise nature of the interaction between the dark sectors is yet unknown, the possibility for the cosmological models permitting nonlinear interaction is equally acceptable to understand the dynamics of the universe. The nonlinear interaction with holographic DE was discussed in the references\cite{Oliveros:2014kla,Bolotin_2015,Arevalo:2011hh}. The authors in reference\cite{Feng:2007wn, Sadri:2019yqs} test the viability of the interacting holographic DE model by using the combined observational data. The present work thus examines the cosmological dynamics in the presence of various nonlinear interaction rates between DM and DE in a spatially flat FLRW universe. 
	
	In this work, we have explored a model of HRDE as a running vacuum by accounting for a nonlinear interaction between
	the DM and the DE. The nonlinear interaction of the form $Q=3bH\rho^{m+s}\rho_m^n\rho_{de}^{-s-n}$
	where b is the coupling constant, powers s,n, and m specify interactions, generalizing several models studied so far in the literature\cite{Arevalo:2011hh}. The models corresponding to various combinations of m, n, and s were found to be different from that of the standard $\Lambda$CDM model\cite{Bolotin_2015}. In particular, we have considered three distinct nonlinear interaction forms for the corresponding values of m,n,s as (m,n,s)=(1,2,-2), (m,n,s)=(1,0,-2), (m,n,s)=(1,1,-2) in our work which gives analytically feasible solutions. For each of these cases, we analytically solve for the Hubble parameter. To examine the observational acceptance of the models, we perform global fittings of the models using the latest observational data SN1a(Pantheon)+CMB(Planck2018)+BAO(SDSS) data set. We have extracted the value of model parameter $\beta$, interaction term $b$, $\Omega_{de0}$, $H_0$, and $M$ for all the three interaction cases using the $\chi^2$ minimization technique. The parameters of all the three cases are found to behave similar values as $\beta \sim 0.45$, $\Omega_{de0} \sim 0.72$, $H_0\sim 68.7$, $M\sim 19.39.$  The value of the interaction term b for NLI1 and NLI3 is $0.003$ whereas, for NLI2, its value is $0.001.$ The positive value of the interaction parameter confirms the transfer of energy from DE to the DM. The evolution of the Hubble parameter using these constraints of model parameter for three interaction cases predicts the presence of the big bang at the origin.
	
	We also derive the deceleration parameter for the three cases of nonlinear interaction. The present value of the deceleration parameter, $q_0$, agrees with the observational results. The evolution of the q-z plot for the three cases predicts a transition from decelerated to the accelerated epoch at the redshift, $z_T\sim 0.72$, which agrees with the range of $\Lambda$CDM model. 
	
	The state finder diagnosis of the three cases shows that this model is distinguishably different from the $\Lambda$CDM model. The trajectories of $\{r,s\}$ parameter of NLI1 and NLI3 are approaching the $\Lambda$CDM fixed point in the future, whereas that of NLI2 is diverging from the $\Lambda$CDM fixed point. The trajectories of three cases lie in the region $r<1$, and $s>0$ resembles the nature of quintessence DE.  
	
	The phase space analysis of the three cases was analyzed. The critical points obtained having both positive and negative eigenvalues represent the prior unstable matter-dominated phase. The critical points having both negative eigenvalues correspond to the stable accelerated epoch. For the case NLI1: the critical point $(0.997,0.003)$ represents an unstable matter-dominated phase, and the critical point $(0,1)$ gives a stable de Sitter phase. For NLI2, the critical point $(1,0)$ corresponds to an unstable matter-dominated epoch, whereas the critical point $(0.001,0.999)$ shows an unstable DE-dominated epoch. For NLI3, the critical point $(1,0)$ represents an unstable matter-dominated early phase, and the critical point $(0,1)$ corresponds to the stable DE-dominated late phase.
	Our analysis shows that all the three interacting cases predict a prior matter-dominated universe with an unstable node and a stable DE-dominated universe with acceleration, similar to the de Sitter phase.  
	
	The authors in the reference\cite{De-Santiago:2016oeu} interpret the  dark sector interaction model in which the coupling form accounts for the interaction between
	the DM and the DE
	$Q=\frac{\acute{\rho}}{3(\alpha+\beta \rho)}$ as a particular case of a running vacuum model of the type $\Lambda(H)=n_0+n_1H^2+n_2H^4$ in which VE decays to DM. In the reference\cite{Bouhmadi-Lopez:2016dcs} authors show that the nonlinear interaction between dark sectors results in an ISW(Integrated Sachs–Wolfe) effect. A more quantitative analysis of these interesting issues of non-linearly interacting models will be the subject of future work.
	
	%\section{Acknowledgment}
	\bibliographystyle{cpc}
	\bibliography{work4reference}
\end{document}